\newcommand{\be}{\begin{equation}}
\newcommand{\ee}{\end{equation}}
\newcommand{\bea}{\begin{eqnarray}}
\newcommand{\eea}{\end{eqnarray}}

\newcommand{\balpha}{\mbox{\boldmath{$\alpha$}}}

\newcommand{\bone}{\mbox{\boldmath{$1$}}}
\newcommand{\bGamma}{\mbox{\boldmath{$\Gamma$}}}
\newcommand{\bbeta}{\mbox{\boldmath{$\beta$}}}

\documentclass{cimento}

\usepackage{graphicx}  
\title{Three-body Casimir-Polder interactions}
\author{Kimball A. Milton\from{ins:ou}, Elom Abalo\from{ins:ou},
Prachi Parashar\from{ins:ou},
    \atque
K. V. Shajesh\from{ins:isu}}
\instlist{\inst{ins:ou} Homer L. Dodge Department of Physics and Astronomy,
University of Oklahoma, Norman, OK 73019, USA
  \inst{ins:isu} 
Department of Materials Science and Engineering, Iowa State University of Science and 
Technology, Ames, Iowa 50011,
USA}
\PACSes{\PACSit{03.70.+k}{11.80.La -- 34.35.+a -- 42.50.Lc}}

\begin{document}

\maketitle

\begin{abstract}
As part of our program to develop the description of three-body effects
in quantum vacuum phenomena, we study the three-body interaction of
two anisotropically polarizable atoms with a perfect electrically
conducting plate, a generalization of
earlier work. 
Three- and four-scattering
effects are important, and lead to nonmonotonic behavior.
\end{abstract}

\section{Introduction}
Even before the discovery of the Casimir effect \cite{casimir},
Casimir and Polder studied the retarded dispersion force between
two atoms, and between one atom and a perfectly  conducting surface \cite{cp}.
The Casimir-Polder force between an atom and a conducting plate
was, until relatively recently, only observed in one experiment
\cite{sukenik}, although the temperature dependence of this force
has now been confirmed \cite{cornell}. In contrast, the experimental
observation of the Casimir-Polder force between atoms has remained
beyond reach. For an elementary review of
aspects of Casimir-Polder forces see Ref.~\cite{miltonrl}.

In this paper we wish to consider three-body Casimir-Polder energies, involving
two polarizable atoms and a perfectly conducting plate.  This was considered
recently, but in a somewhat restricted geometry \cite{lopez}.  A scalar
analog was also recently examined by Shajesh and Schaden \cite{shajesh}.
(In the nonretarded regime, this was also considered by de Melo e Souza
{\it et al.}, following earlier work described in Ref.~\cite{souza}.)
Here we wish to examine the situation a bit more generally.  Our interest
is not so much in finding observable effects, which are probably difficult to
observe, but to understand the general features of three-body interactions; 
this is a small part of our continuing efforts in this direction.

\section{Two-body energy}
For simplicity, we assume we are in the fully retarded regime, that is,
the atoms are sufficiently far from each other or the  plate
so that we can describe them by their static electric polarizabilities,
\be
\balpha_a=\balpha_a(\omega=0), \quad a=1, 2. \label{staticpol}
\ee
We also neglect magnetic polarizability, and quadrupole and higher
multipole moments of the atoms.
We will, however, assume that the atoms are not isotropically polarizable.
In the isotropic case, the two-body interaction energy between the atoms,
a distance $r$ apart, is given by the famous
formula ($\hbar=c=1$)
\be
E_{12}=-\frac{23}{4\pi}\frac{\alpha_1\alpha_2}{r^7},\label{cpatoms}
\ee
but it is considerably more complicated when the atoms are not isotropically
polarizable:
\be
E_{12}=-\frac{13\, \mbox{Tr}\, 
(\balpha_1\cdot \balpha_2)
-56\, (\mathbf{\hat r}\cdot\balpha_1\cdot\balpha_2\cdot\mathbf{\hat r})
+63\, (\mathbf{\hat r}\cdot\balpha_1\cdot \mathbf{\hat r})
(\mathbf{\hat r}\cdot\balpha_2\cdot \mathbf{\hat r})}{8\pi r^7},
\ee
where $\mathbf{r}=r\mathbf{\hat r}$ is the relative position vector
of the two atoms.  This formula was apparently first given by
Craig and Power \cite{craig}.
In contrast, the interaction energy of a polarizable atom with a
perfectly conducting plate a distance $Z$ away is very simple,
\be
E_{\rm CP}=-\frac{\mbox{Tr}\,\balpha}{8\pi Z^4}.\label{dpatomwall}
\ee
(One might recall that the dimension of $\balpha$ is $(\mbox{length})^3$,
and that $\balpha$ is necessarily a symmetric matrix.)

We now wish to consider what happens when all three bodies are present,
the two atoms and the perfectly conducting plate.

\section{Three-body energy}
We wish to consider two polarizable atoms in the neighborhood of a 
perfectly conducting plate.  
\begin{figure}
\centering
\includegraphics{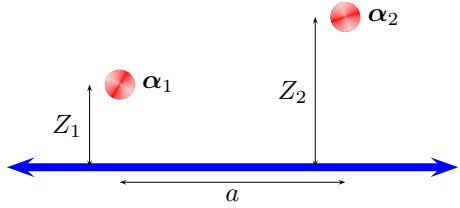}
\caption{\label{3bodycp} Two anisotropically polarizable atoms 
interacting with a perfectly conducting
plate.  The atom $a$, $a=1,2$,  has polarizability $\balpha_a$, and is a 
distance $Z_a$ above
the plate.  The distance between the atoms in the
direction parallel to the plate is $a$.  The difference in heights above
the plate is $\Delta Z=Z_1-Z_2$.}
\end{figure}
The geometry is sketched in Fig.~\ref{3bodycp}.  The two atoms have 
polarizabilities $\balpha_a$, and are located at distances from the plate 
$Z_a$, $a=1, 2$.  The difference in
heights from the plate is $\Delta Z=Z_1-Z_2$, and the horizontal distance 
(parallel to the plate)  between the two atoms is called $a$.

To compute the interaction energy, 
we use the multiple scattering formalism.  For the case
at hand, where only a single scattering with each atom need be considered, the
two-body terms are (this is just the famous TGTG formula in weak coupling
\cite{Milton:2008vr,yale})
\be
E_{ab}=\frac{i}2\int \frac{d\omega}{2\pi}
\mbox{Tr}\,\mathbf{\tilde T}_a\mathbf{\tilde T}_b,
\ee
where $a,b=1,2,3$, and we denote the two atoms as bodies 1 and 2, while
the perfectly conducting plate is body 3. Here the modified scattering operator is
\be
\mathbf{\tilde T}=\mathbf{T}\bGamma_0=\mathbf{V}(\mathbf{1}-\bGamma_0\mathbf{V}
)^{-1}\bGamma_0. 
\ee
The free propagator $\bGamma_0$ is \cite{brevikvol}
\be
\bGamma_0(\mathbf{r,r'})=-\frac{e^{-|\zeta| R}}{4\pi R^3}\left[\bone u(|\zeta|R)
-\mathbf{\hat R\hat R}v(|\zeta|R)\right],\quad \mathbf{R=r-r'},
\ee
where $\zeta=-i\omega$ is the imaginary frequency, and
\be
u(x)=1+x+x^2,\quad v(x)=3+3x +x^2.
\ee
The scattering matrix for the atom is just the potential,
\be
\mathbf{T}_{1,2}=\mathbf{V}_{1,2}=4\pi\balpha_{1,2}\delta(\mathbf{r-r}_{1,2}),
\ee
where $\mathbf{r}_a$ are the positions of the atoms.
The structure in which the scattering matrix for the plate appears is
\be
\bGamma_0 \mathbf{T}_3\bGamma_0=\bGamma_3-\bGamma_0;
\ee
$\bGamma_3$ is the Green's dyadic for the plate.
In fact, with $\mathbf{\hat n}$ denoting the normal to the plate, 
\be
(\bGamma_3-\bGamma_0)(\mathbf{r,r'})=-\bGamma_0(\mathbf{r,r'}-
2\mathbf{\hat n}\mathbf{\hat n\cdot r}')\cdot
(\bone-2\mathbf{\hat n\hat n})\equiv -\overline{\bGamma}(\mathbf{r,r'}).
\ee
This is just the image construction for a perfect electric mirror 
\cite{levine}.

Using these constructions, the results in the previous section are easily
derived.  Our concern here is with the three-body terms.  These are of
three, related, types \cite{shajesh11}, where the subscripts on the right-hand
side refer to the ordering of the scattering between the bodies:
\be
\Delta E_3=E_{123}+E_{213}+E_{1323}.
\ee
The three-scattering energy is
\be
E_{123}=\frac{i}2\int\frac{d\omega}{2\pi}
\mbox{Tr}\,\mathbf{\tilde T}_1\mathbf{\tilde T}_2\mathbf{\tilde T}_3
=\frac{1}2\int_{-\infty}^\infty\frac{d\zeta}{2\pi}\mbox{Tr}\, \mathbf{V}_1
\bGamma_0\mathbf{V}_2\overline{\bGamma},
\ee
while $E_{213}$ is obtained by interchanging 1 and 2.
The four-scattering contribution is
\be
E_{1323}=\frac{i}2\int\frac{d\omega}{2\pi}
\mbox{Tr}\,\mathbf{\tilde T}_1 \mathbf{\tilde T}_3 
\mathbf{\tilde T}_2 \mathbf{\tilde T}_3=-\frac12\int_{-\infty}^\infty
\frac{d\zeta}{2\pi}
\mbox{Tr\,}\mathbf{V}_1 \overline{\bGamma} \mathbf{V}_2 \overline{\bGamma}.
\ee

It is very straightforward to work out these contributions in the static
approximation (\ref{staticpol}), where we ignore the frequency dependence of the
polarizability. Let us now choose the plate to lie in the $x$-$y$ plane.
Because of the  operator appearing in 
$\overline{\bGamma}$, it is convenient to define $\bbeta_a=
(\bone_\perp-\mathbf{\hat z\hat z})\cdot\balpha_a$, $\bone_\perp=\mathbf{
\hat x\hat x+\hat y\hat y}$.  We can write all the energy 
contributions, including the two-atom ones, as follows:
\begin{eqnletter}
E_{12}(\mathbf{r}_{12})&=&-F(\mathbf{r}_{12},\mathbf{r}_{21};\balpha_1,
\balpha_2),\\
E_{123}(\mathbf{r}_{12},\mathbf{r}_{2\bar1})&=&F(\mathbf{r}_{12},\mathbf{r}_{2
\bar 1};\balpha_2,\bbeta_1),\\
E_{213}(\mathbf{r}_{21},\mathbf{r}_{1\bar2})&=&F(\mathbf{r}_{21},\mathbf{r}_{1
\bar 2};\balpha_1,\bbeta_2),\\
E_{1323}(\mathbf{r}_{2\bar1},\mathbf{r}_{1\bar2})&=&
-F(\mathbf{r}_{2\bar1},\mathbf{r}_{1\bar 2};\bbeta_1,\bbeta_2).
\end{eqnletter}
Here $\mathbf{r}_{ab}=\mathbf{r}_a-\mathbf{r}_b$, and the positions of the
images of the atoms are denoted by $\mathbf{r}_{\bar{a}}$.
These relative position vectors satisfy
\begin{eqnletter}
\mathbf{r}_{21}+\mathbf{r}_{1\bar 2}+\mathbf{r}_{\bar 2\bar1}
+\mathbf{r}_{\bar12}&=&\mathbf{0},\\
(\mathbf{r}_{21}+\mathbf{r}_{1\bar2})\cdot(\mathbf{r}_{1\bar 2}+\mathbf{r}_{
\bar12})&=&0.
\end{eqnletter}
The function appearing in the energies is
\bea
F(\mathbf{x,y};\balpha,\bbeta)&=&\frac1{4\pi x^3y^3(x+y)}
[A(x,y)\mbox{Tr}\,
(\balpha\cdot\bbeta)-
B(x,y)(\mathbf{\hat y\cdot\bbeta\cdot\balpha\cdot \hat y})\\
&&\qquad\mbox{}-B(y,x)
(\mathbf{\hat x\cdot\balpha\cdot\bbeta\cdot \hat x})+C(x,y)(\mathbf{
\hat x \cdot\balpha\cdot \hat y})(\mathbf{\hat y\cdot \bbeta\cdot \hat x})
],\nonumber
\eea
where
\begin{eqnletter}
A(x,y)&=&\frac8{(x+y)^4}\left[x^4+5x^3y+14x^2y^2+5xy^3+y^4\right]
\stackrel{x=y}{\longrightarrow}13,\\
B(x,y)&=&\frac8{(x+y)^4}\left[3x^4+15x^3y+26x^2y^2+10xy^3+2y^4\right]
\stackrel{x=y}{\longrightarrow}28,\\
C(x,y)&=&\frac{48}{(x+y)^4}\left[x^4+5x^3y+9x^2y^2+5xy^3+y^4\right]
\stackrel{x=y}{\longrightarrow} 63.
\end{eqnletter}

\section{Special Cases}
\subsection{Isotropically polarizable atoms, equidistant from the plate}
\begin{figure}
\centering
\includegraphics{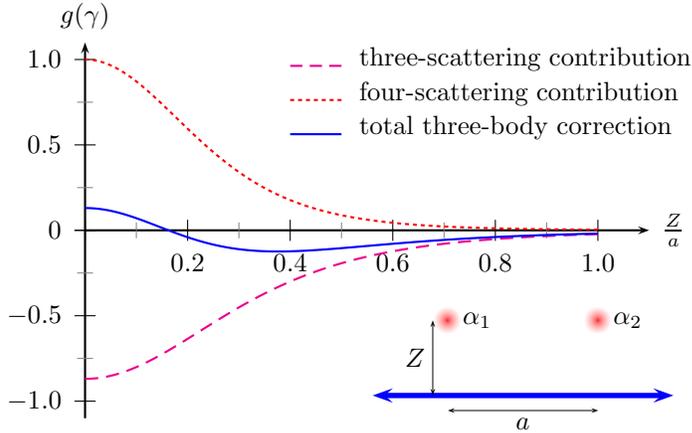}
\caption{\label{fig:g} The three-body correction to the Casimir-Polder
force between two atoms in the presence of a perfectly conducting plate.  Here
it is assumed that the atoms are isotropically polarizable, 
and are equidistant from the plate. 
 What is plotted is the ratio of the three-body correction
relative to the two-atom energy, plotted as a function of $Z/a$. 
The variable $\gamma$ is defined in Eq.~(\ref{gamma}).}
\end{figure}

Let us first consider isotropic atoms equidistant from the plate,
so $\Delta Z=0$ and $Z_1=Z_2=Z$.  Defining
\be
\gamma=\sqrt{1+\frac{4 Z^2}{a^2}},\label{gamma}
\ee
we express the three-body interaction in terms of the two-atom Casimir-Polder
 energy,
\be
\Delta E_3=E_{123}+E_{213}+E_{1323}=g(\gamma) E_{12},
\ee
where in this case $E_{12}$ is given by Eq.~(\ref{cpatoms}).
Here
\be
g(\gamma)=-\frac{64(1+4\gamma)}{23\gamma^3(1+\gamma)^4}
+\frac1{\gamma^{7}},\label{isog}
\ee
where the last term is the four-scattering contribution.
This function vanishes as $\gamma\to\infty$, that is, at large
distances from the plate, and has
a maximum negative deviation of about 12\%,
and has the value on the plate, at $Z=0$, of $g(1)=3/23\approx 0.13$,
as shown in Fig.~\ref{fig:g}. (In the scalar analog considered
in Ref.~\cite{shajesh}, the correction on the plate is $g(1)=-1$, so the
three-body energy cancels the two-atom energy when the atoms touch the plate.\footnote{$g(\gamma)$
for the scalar analog reported in Ref.~\cite{shajesh} has an error of
a factor of 2 in its first term.})
Note that the three-scattering terms,
which are always negative,  dominate for $Z/a>0.16$, 
but very close to the plate the four-scattering term,
which is always positive, causes
the sign of the correction to reverse. However, close to the plate,
these corrections are negligible compared to the two-body atom-wall energy
(\ref{dpatomwall}), because 
\be
\frac{E_{12}}{E_{\rm CP}}=\frac{\alpha_2}{r^3}\frac{46}3\left(\frac{Z}{r}
\right)^4\ll1,
\ee
since typical values of $\alpha\sim (10^{-8}\,\mbox{cm})^3$, while
the separation consistent with our macroscopic approximation is not
going to be smaller than $r\sim 10^{-6}$ cm.
 We might note that if the
atoms and the plate are all equidistant, $Z=a$, so the three-body 
effect is small, and the four-scattering contribution is totally
negligible.

\subsection{Anisotropically polarizable atoms, equidistant from the plate}
As a second example, we show the result considered in Ref.~\cite{lopez},
when the atoms are at equal heights, $\Delta Z=0$, 
with a simple anisotropic polarizability:
\be
\balpha_a=\mbox{diag}(\alpha_\perp^a,\alpha_\perp^a,\alpha_z^a),\quad a=1,2.
\ee
Then 
\bea
E_{123}&=&\frac{2}{\pi a^7}\frac1{\gamma^{5}}\frac1{(1+\gamma)^{5}}
[\alpha_z^1\alpha_z^2
(-3-15\gamma-24\gamma^2+10\gamma^4+5\gamma^5+\gamma^6)\\
&&\quad\mbox{}+\alpha_\perp^1\alpha_\perp^2(3+15\gamma+28\gamma^2+20\gamma^3
+6\gamma^4-5\gamma^5-\gamma^6)].\nonumber
\eea
This agrees with the result given in Ref.~\cite{lopez}, when the identical
contribution of $E_{213}$ is included.

The four-scattering term may be given in a similar form
\bea
E_{1323}&=&-\frac1{8\pi a^7\gamma^{11}}[\alpha_z^1\alpha_z^2(63-70\gamma^2
+20\gamma^4)\\
&&\quad\mbox{}+\alpha^1_\perp\alpha^2_\perp(63-56\gamma^2+26\gamma^4)
+(\alpha_\perp^1\alpha_z^2+\alpha_z^1\alpha_\perp^2)63(\gamma^2-1)].\nonumber
\eea
This is identical to the corresponding term in Ref.~\cite{lopez}. 
These energies, of course, agree with the result (\ref{isog}) in the
isotropic case $\alpha^a_\perp=\alpha^a_z$.

\begin{figure}
\centering
\includegraphics{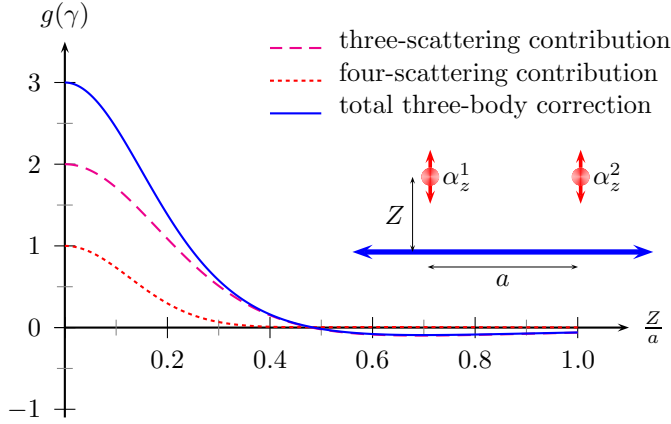}
\caption{\label{fig:gzz} The three-body correction to the Casimir-Polder
force between two atoms in the presence of a perfectly conducting plate.  Here
it is assumed that the atoms are completely anisotropic,
being polarizable only in the direction perpendicular to the plate, and are 
the same distance
from the plate.  What is plotted is the ratio of the three-body correction
relative to the two-atom energy, plotted as a function $Z/a$. The variable
$\gamma$ is again defined in Eq.~(\ref{gamma}).}
\end{figure}

 We plot the contributions for the $\alpha_z^1\alpha_z^2$ terms in 
Fig.~\ref{fig:gzz}.
That is, suppose the atoms are only polarizable in the $z$ direction.  Then
the two-body Casimir-Polder interaction between the atoms is
\be
E_{12}=-\frac{13}{8\pi a^7}\alpha^1_z\alpha^2_z,
\ee
and we normalize the three- and four-scattering contributions relative
to this value:
\be
\frac{2E_{123}}{E_{12}}=g_3(\gamma), \quad \frac{E_{1323}}{E_{12}}=
g_4(\gamma),\quad
g(\gamma)=g_3(\gamma)+g_4(\gamma).
\ee
Again it is seen that the four-scattering contribution is significant only
very close to the plate.  Now the sign of the correction is negative only 
for $Z/a>0.485$.

\subsection{Anisotropically polarizable atoms, unequal distances from the plate}
\begin{figure}
  \begin{center}
\includegraphics{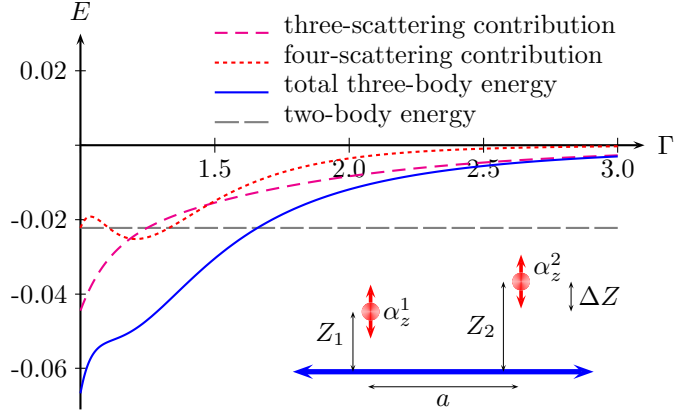}
    \caption{\label{es75} The three-body correction to the Casimir-Polder
energy (in units of $\alpha_z^1\alpha_z^2/r^7$)
 between two atoms in the presence of a perfectly conducting plate.  Here, 
it is assumed that the atoms are
polarizable only in the direction perpendicular to the plate, and
are at different distances from the 
plate. 
 Note that although for large distances, that is, for large values of
$\Gamma$, the three-scattering contribution dominates, for values close to
$\Gamma=1$ the four-scattering contribution is significant, and leads to
nonmonotonicity in the slope.  The energies plotted are for $a/r=0.75$,
or $\Delta Z/a\approx0.88$,
where the effect of the four-scattering term is most significant.}
  \end{center}
\end{figure}

As a final example, let us  assume that the atoms are only
polarizable in the $z$ direction,  $\alpha^a_z\ne 0$, $\alpha^a_\perp=0$,
but that the atoms are at different distances from the plate, $\Delta Z\ne0$.
Then, with the distances between the atoms being $r$, and the distance
between one atom and the image of the other being $R\equiv\Gamma r$,
that is,
\begin{equation}
R^2=a^2+(Z_1+Z_2)^2,\quad
r^2=a^2+(Z_1-Z_2)^2,
\end{equation}
we easily find
\begin{eqnletter}
E_{12}&=&-\frac{\alpha^1_z\alpha^2_z}{8\pi r^7}\left(20-70\frac{a^2}{r^2}
+63\frac{a^4}{r^4}\right),\\
E_{123}=E_{213}&=&-\frac{2\alpha^1_z\alpha^2_z}{\pi r^7\Gamma^5(1+\Gamma)^5}
\bigg[2\Gamma^2(1+\Gamma)^2(1+3\Gamma+\Gamma^2)\\
&&\quad\mbox{}-\frac{a^2}{r^2}(1+\Gamma)^2(3+9\Gamma+11\Gamma^2+9\Gamma^3+3
\Gamma^4)\nonumber\\
&&\quad\mbox{}+6\frac{a^4}{r^4}
(1+5\Gamma+9\Gamma^2+5\Gamma^3+\Gamma^4)\bigg],\nonumber\\
E_{1323}&=&-\frac{\alpha^1_z\alpha^2_z}{8\pi r^7\Gamma^{7}}\left(20
-70\frac{a^2}{\Gamma^2r^2}+63\frac{a^4}{\Gamma^4r^4}\right).
\end{eqnletter}
We plot these functions versus $\Gamma$ in Fig.~\ref{es75}. 

The figure shows that although the four-scattering term is generally negligible
compared to the three-scattering term, it is significant near the plate.
In fact, for $\Gamma=1$, which corresponds to one atom touching the plate,
the following relation holds:
\be
E_{123}=E_{1323}=E_{12}, \quad r=R,
\ee
so the total energy is (again, we exclude the infinite two-body atom-wall energy)
\be
E=E_{12}+2E_{123}+E_{1323}=4E_{12}
\ee
at that unphysical point.
Nonmonotonic effects can be seen when $a/r$ is sufficiently near 1, or
$\Delta Z/a$ sufficiently small.

To close this section, we consider the same geometry, but atoms that are
only polarizable transversely, $\alpha^a_z=0$, $\alpha^a_\perp\ne0$.
Then we immediately find
\begin{eqnletter}
E_{12}&=&-\frac{\alpha_\perp^1\alpha_\perp^2}{8\pi r^7}\left(26-56\frac{a^2}
{r^2}+63\frac{a^4}{r^4}\right),\\
E_{123}=E_{213}&=&\frac2{\pi r^7\Gamma^5(1+\Gamma)^5}\big[2\Gamma^2(1
+5\Gamma+14\Gamma^2+5\Gamma^3+\Gamma^4)\\
&&\mbox{}-\frac{a^2}{r^2}(3+15\Gamma+28\Gamma^2+20\Gamma^3+28\Gamma^4
+15\Gamma^5+3\Gamma^6)\nonumber\\
&&\mbox{}+6\frac{a^4}{r^4}(1+5\Gamma+9\Gamma^2+5\Gamma^3+\Gamma^4)\big],
\nonumber\\
E_{1323}&=&-\frac{\alpha_\perp^1\alpha_\perp^2}{8\pi r^7\Gamma^{7}}
\left[26-56\frac{a^2}{\Gamma^2r^2}+63\frac{a^4}{\Gamma^4r^4}\right].
\end{eqnletter}
\begin{figure}
  \begin{center}
\includegraphics{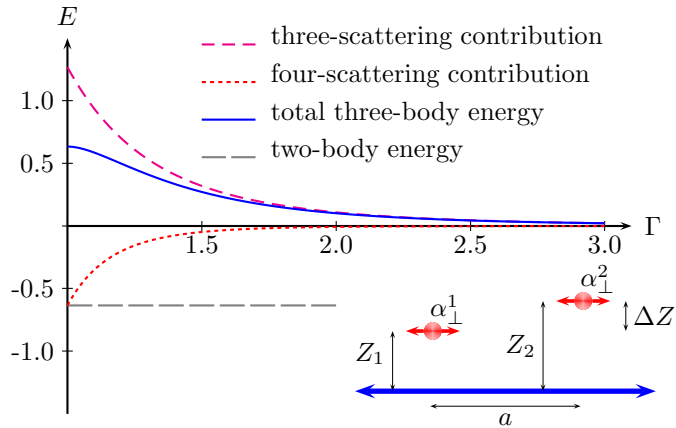}
   \caption{\label{eperp} Here we plot the three-body correction
(in units of $\alpha_\perp^1\alpha_\perp^2/r^7$) to the
Casimir-Polder energy $E_{12}$ 
for $a/r=0.5$, for atoms polarizable only in directions
parallel to the plate.  
Now there are no nonmonotonic
effects, and the three-scattering contribution is dominant, although the
four-scattering term can be significant near the plate.}
  \end{center}
\end{figure}

Now the effects of the four-scattering terms are always much less dramatic.
For example, for $a/r=0.5$, in Fig.~\ref{eperp} we plot the contributions
to the correction to the Casimir-Polder energy between the atoms.
The three-scattering contribution always dominates, although the 
four-scattering term can be important near the plate.  Now, at the
unphysical point $\Gamma=1$ we have
\be
E_{123}=-E_{1323}=-E_{12},
\ee
so the total energy is zero there. Thus, $g(1)=-1$ for this case,
similar to the scalar analog considered in Ref.~\cite{shajesh},
where the three-body energy cancels the two-body energy when the
atoms touch the plate. Of course, this result
does not  include the infinite
Casimir-Polder energy (\ref{dpatomwall}) corresponding to the atom touching the plate.

\section{Conclusions}
We have illustrated, in the simple context of two polarizable atoms
interacting with each other and a nearby perfectly conducting plate, the three-body
interactions beyond the two-body Casimir-Polder forces.  These three-body
energies break up into three- and four-scattering terms in this context,
where the atom interactions are regarded as weak.  This work generalizes
that of Ref.~\cite{lopez}, which considered atoms equidistant from the plate,
and that of Ref.~\cite{shajesh}, which considered a scalar analog
of the Casimir-Polder interaction.  We observe regions of nonmonotonicity in
the energy, which is reminescent of results found earlier, numerically,
in the context of rectangular objects near conducting surfaces 
\cite{zaheer,rodriguez}.

These effects are of conceptual interest only, because they generally
represent small corrections to the Casimir-Polder interaction
between atoms, which itself has never been directly observed, due to
lack of sufficiently refined experimental technique. 
Where the three-body interactions are comparable to the two-body atom-atom
energy, they are dwarfed by the two-body atom-wall interaction 
(\ref{dpatomwall}).  The
importance of this work lies in its contribution to our developing
understanding of three-body effects in Casimir or quantum vacuum energy
calculations.  Further examples of these effects, in more nontrivial
contexts, will appear elsewhere.

\acknowledgments
This work was supported in part by grants from the US National Science 
Foundation, the US Department of Energy, and the Julian Schwinger Foundation.  
We thank Fardin Kheirandish and Reinaldo de Melo e Souza for discussions.
KVS would like to thank Martin Schaden for contributions in the early
stages of this work, and Tom Sizmur for discussions.

\end{document}